\documentclass{elsarticle}
\usepackage{multirow}
\usepackage{amsmath,amssymb,amsfonts}
\usepackage{graphicx}
\usepackage{subcaption}
\usepackage{url}
\usepackage{caption}
\begin{document}  

\title{Toward a Deep Learning-Driven Intrusion Detection Approach for Internet of Things}

\author[label1]{Mengmeng Ge}
\author[label1]{Naeem Firdous Syed}
\author[label2]{Xiping Fu}
\author[label1]{Zubair Baig}
\author[label1]{Antonio Robles-Kelly}

\address[label1]{(mengmeng.ge, naeem.syed, zubair.baig, antonio.robles-kelly)@deakin.edu.au, School of Information Technology, Deakin University, Geelong, VIC, Australia}
\address[label2]{fxpfxp0607@gmail.com, PredictHQ Limited, Auckland, New Zealand}

\begin{abstract}

 Internet of Things (IoT) has brought along immense benefits to our daily lives encompassing a diverse range of application domains that we regularly interact with, ranging from healthcare automation to transport and smart environments. However, due to the limitation of constrained resources and computational capabilities, IoT networks are prone to various cyber attacks. Thus, defending the IoT network against adversarial attacks is of vital importance. In this paper, we present a novel intrusion detection approach for IoT networks through the application of a deep learning technique. We adopt a cutting-edge IoT dataset comprising IoT traces and realistic attack traffic, including denial of service, distributed denial of service, reconnaissance and information theft attacks. We utilise the header field information in individual packets as generic features to capture general network behaviours, and develop a feed-forward neural networks model with embedding layers (to encode high-dimensional categorical features) for multi-class classification. The concept of transfer learning is subsequently adopted to encode high-dimensional categorical features to build a binary classifier. Results obtained through the evaluation of the proposed approach demonstrate a high classification accuracy for both binary and multi-class classifiers. 

\end{abstract}

\maketitle

\section{Introduction}

Internet of Things (IoT) has proliferated into our daily life at a very rapid pace. Recent statistics show IoT devices have pervaded into various domains, including smart cities (28.6\%), industrial IoT (26.4\%), eHealth (22\%), smart homes (15.4\%) and smart vehicles (7.7\%)~\cite{IoTStats}. Moreover, the number of IoT devices in use has seen a steep rise from 15.4B in 2015 to 26.7B in 2019, with numbers continuing to climb as households and businesses alike are increasingly becoming dependent on Internet-based services for their routine activities. Many critical infrastructures also comprise of IoT devices. For instance, the smart grid has been constantly interacting with IoT devices that measure electricity grid parameters and report these back to a central computing device, for processing. 

Advances in IoT technologies have demonstrated benefits to users. However, IoT devices have also been exposed to cyber threats as posed by the adversary due to the heterogeneity in the communication protocols and constrained resources and large-scale proliferation of IoT devices. For example, one commonly used application protocol in IoT networks is the Message Queuing Telemetry Transport (MQTT). The MQTT protocol is lightweight in nature and based upon the publish-subscribe model of communication; using the MQTT protocol, IoT devices can communicate with centralised computing nodes through intermediary brokers \cite{MQTT}. However, the MQTT protocol suffers from many vulnerabilities that open up the IoT space to the ever-evolving adversarial threat, including device compromise, data theft, limited resource access for legitimate IoT nodes, and Man-in-The-Middle (MiTM) attacks for communicating IoT device pairs.

Intrusion detection systems are designed and developed to foster rapid and accurate identification of malicious attempts by the adversary to penetrate into IoT networks and to cause system disruption. The application of artificial intelligence (AI) for the timely detection of malicious attacks, is both necessary as well as effective. The range of AI techniques applicable for network intrusion detection (i.e., for detecting malicious attacks through analysis of network traffic) is very broad. Many AI techniques, including support vector machines (SVM) \cite{IoTSVM}, Bayesian networks~\cite{IoTBayesian}, principal component analysis~\cite{IoTPCA}, and genetic algorithms~\cite{IoTGA}, have been popularly deployed for the detection of anomalous behaviour in IoT networks. However, deep learning based intrusion detection for IoT networks is still under-researched. 

In this work, we propose a novel intrusion detection approach against cyber attacks for IoT networks based upon the concept of deep learning. We design the classifier as a feed-forward neural networks model with embedding layers to identify four categories of attacks, namely distributed denial of service (DDoS), denial of service (DoS), reconnaissance, and information theft, whilst differentiating these from the legitimate network traffic. In addition, the encoding of high-dimensional categorical features is extracted through network embedding and applied to binary classifier via transfer learning. 

A preliminary version of this paper appeared in~\cite{Ge2019PRDC}. We have extended the earlier version with (1) a thorough investigation of header fields and inclusion of new features, (2) design of a feed-forward neural networks model with embedding to deal with high-dimensional categorical features for multi-class classification, and (3) employment of transfer learning to build a feed-forward neural networks model for binary classification with low-dimensional continuous vector representation of high-dimensional categorical features. The proposed classifiers in this extended work have demonstrated a significant improvement of performance regarding all evaluation measures. Main contributions of the paper are as follows.
\begin{itemize}
\item Adoption of a newly published IoT dataset with realistic attack traffic and simulated IoT traffic in a smart home scenario;
\item Generation of generic features based on the header field information in individual IP packets to avoid the attack-oriented feature selection procedure;
\item Design and development of the deep learning based intrusion detection approach including both binary and multi-class classifications for IoT networks with high accuracy.
\end{itemize} 

The rest of the paper is organised as follows. Section 2 provides the background to the study. Our proposed intrusion detection framework is presented in Section 3. In Section 4, we provide a description of the dataset, design of the experimental setup, and analysis of the obtained results. A discussion on future directions of the work is provided in Section 5, and the paper is concluded in Section 6. 

\section{Related Work}

In this section, we briefly survey the state-of-the-art literature in intrusion detection approaches for traditional and IoT networks with the support of machine learning and deep learning techniques.

\textbf{Approaches for traditional networks:} Haddadi et al. \cite{feedForwardIDS1} proposed a feed forward neural network-based approach to detect DoS, root to local (R2L), user to root (U2R), and probe attacks in the DARPA dataset \cite{McHugh2000}. A feed forward neural network comprises an input, a hidden and an output layer, with each layer consisting of several neurons. During the training phase, various strategies and mathematical operations are incorporated into the neural network to tune weights that interconnect the three layers until an optimal solution is obtained. The back propagation algorithm is the most commonly adopted algorithm in training neural networks to update and choose optimal weights of the interconnections which minimise the loss. Authors reported an accuracy of over 96\% in detecting three attack categories among four categories available in the DARPA dataset.

An approach using an ensemble of classifiers was proposed in \cite{RAJKUMAR20111328} to detect DDoS attacks. In their approach, the dataset was split into smaller subsets and each subset was trained using an ensemble of classifiers. Results of individual classifier were combined through the adoption of a weighted majority voting technique which assigns a different weight to each algorithm. The back propagation algorithm was incorporated into the classifier. Their approach obtained an accuracy of 99.4\% in classifying attacks in several publicly available datasets.

Al-Zewairi et al. \cite{deepLearningUNSW} proposed a deep learning model for detecting network based attacks in the UNSW-NB 15 dataset \cite{Moustafa2016}. Their model consists of five hidden layers along with ten neurons distributed across each hidden layer. A 10-fold cross validation technique was implemented by the authors to train the model on the full dataset in 10 epochs. In addition, the importance of features was ranked using the Gedeon method to identify top contributing features. The proposed approach achieved an accuracy of approximately 99\% on the selected dataset. 

An approach based on the radial basis function was proposed in \cite{DeepRadial} to detect DoS attacks. Authors applied radial basis function as the first layer to reduce the training bias which improves the weight optimisation technique of traditional neural networks. The proposed approach achieved an accuracy of 99.69\% when tested on the NSL KDD \cite{kdd} and the UNSW-NB 15 \cite{Moustafa2016} datasets. In a similar work, Shone et al. \cite{Shone2018} adopted a deep network composed of deep auto encoders stack to perform an unsupervised feature learning which is further combined with a random forest classifier. Their approach was test via the NSL KDD \cite{kdd}.

\textbf{Approaches for IoT networks:} Koroniotis et al. \cite{Koroniotis2018arxiv} deployed an IoT testbed to generate an IoT dataset, namely Bot-IoT, due to the lack of realistic IoT traffic in existing datasets. Authors incorporated instances of legitimate and simulated IoT traffic into the dataset along with the attack traffic, including DDoS, DoS, reconnaissance and information theft. Several new features were developed based on correlation coefficient and joint entropy techniques and three machine learning and deep learning algorithms were adopted. The classifiers were evaluated on 5\% of the original dataset for binary classification (i.e., normal traffic and attack traffic of each subcategory) and achieved good accuracy demonstrated through the results. However, the extracted dataset contains imbalanced normal and attack traffic as the number of instances in some attack categories (i.e., DDoS, DoS, and reconnaissance) is much higher than the number of instances in the normal traffic. 

Abeshu et al. \cite{abeshu2018deep} proposed a stacked auto-encoder-based unsupervised deep learning framework for distributed attack detection in IoT networks with a fog computing layer. In specific, fog nodes are responsible for parallel training and model/parameter update. Authors pre-trained the stacked auto-encoder model with unlabelled training data to extract hidden features which were then applied to test data for classification. They evaluated the proposed approach using the NSL KDD \cite{kdd} with 41 features. The deep learning model obtained a higher detection accuracy compared to a shallow learning model (i.e., without pre-training).

Thamilarasu et al. \cite{thamilarasu2019towards} employed a deep belief network (DBN) to construct a feed-forward deep neural network (DNN) for anomaly-based intrusion detection in IoT networks. In specific, DBN layers can be pre-trained using unsupervised learning technique and then taken as hidden layers in the DNN. Authors implemented an IoT testbed with 6 sensors to form a smart home scenario and evaluated the proposed model using the network traffic generated from the testbed. The model achieved high precision, recall, and F1 score for various attacks compared to an existing intrusion detection approach based on inverse weight clustering. 

Roopak et al. \cite{roopak2019deep} compared the performance of four deep learning models in DDoS attacks against IoT networks. The four models include multi-layer perceptron (MLP), convolution neural network (CNN), long short-term memory (LSTM), and a combination of CNN and LSTM. Authors adopted the CICIDS2017 dataset with 82 flow-based features for evaluation. Comparison results demonstrated that a combination of CNN and LSTM achieved a better performance while MLP had the least performance compared to other neural networks models.

Baig et al. \cite{BAIG2020198} proposed the usage of average dependence estimators for detecting DoS attacks in smart IoT sensors. Authors deployed a custom IoT testbed via MQTT protocol and incorporated DoS attack scenarios based on the protocol. Several packet level features such as source \& destination IP addresses and port numbers, frame length, IP packet length, TCP segment length, and header length, were used in generating the dataset. Two estimators, namely averaged one-dependence estimator (A1DE) and averaged two-dependence estimator (A2DE), were adopted and compared with several traditional machine learning algorithms. Evaluation results indicated that A1DE/A2DE-based classifier achieved higher accuracy compared with other classification algorithms when tested on the custom dataset and BoT-IoT testbed \cite{Koroniotis2018arxiv}. 

\textbf{Summary:} there is little research on deep learning based intrusion detection solutions for IoT networks. Among current approaches, some approaches used the datasets without IoT traces~\cite{abeshu2018deep, roopak2019deep} while others focused on binary classification~\cite{Koroniotis2018arxiv, thamilarasu2019towards}. We can see that one critical reason could be the lack of reliable IoT datasets with the inclusion of attack traffic. In this paper, we adopt the newly published Bot-IoT dataset in \cite{Koroniotis2018arxiv}, develop a novel deep learning based detection approach with generic features in packet level, and perform both binary and multi-class classification. 

\section{Proposed Framework}

\begin{figure*}[h!]
  \centering
 \includegraphics[width=0.95\textwidth]{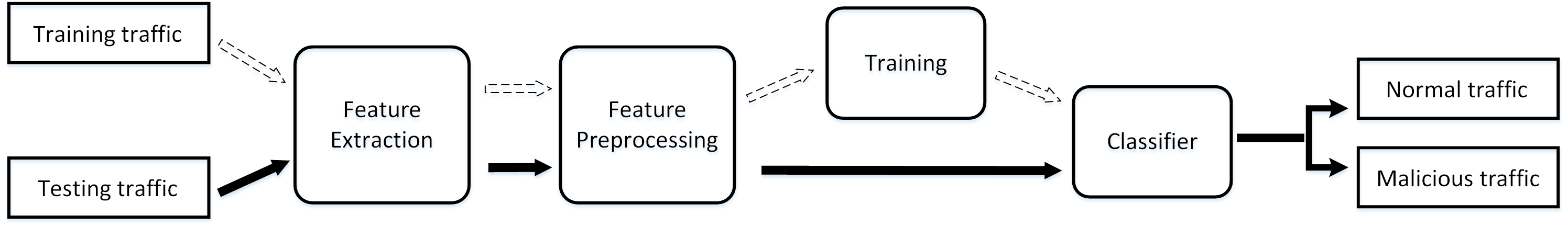}
\caption{Workflow of the framework.}
\label{fig_framework}
\end{figure*}

In this section, we provide a description of the proposed framework for detecting intrusions in IoT networks. Figure~\ref{fig_framework} shows the framework comprising of four phases: feature extraction, feature preprocessing, training and classification. We explain them in detail as follows.

\begin{itemize}
\item \textbf{Feature extraction:} the first phase is to collect the raw network traffic using a network analyser tool (e.g., tcpdump). Relevant fields are then extracted from packets in the raw network traffic, with each field referring to a feature. Features adopted in this work are not based on aggregated packets but header fields from individual packets. Hence, generic features of the traffic could be captured rather than generating attack specific features which might only work for detecting specific attack behaviours. We focus on header fields of an IP packet, such as frame, IPv4/IPv6 and TCP/UDP related information. Details of the extracted features are discussed in Section~\ref{feature_extract}.

\item \textbf{Feature preprocessing:} the second phase is to preprocess the extracted field information by combining, removing, and encoding column values. For categorical data, we apply one-hot encoding to low-dimensional feature columns and embedding (incorporated with the feed-forward neural networks model) to high-dimensional feature columns. The preprocessed data is split into training and testing sets which are used in the training phase and classification phase respectively. Packets in both training and testing sets are labeled as either malicious or normal. Malicious packets are further labelled with the specific attack category it belongs to. Testing packets are used only for evaluating the model accuracy in predicting attacks. Details of the feature preprocessing procedures are discussed in Section~\ref{feature_preprocess}.

\item \textbf{Training:} the training phase takes the processed data from the training set and feeds it into a feed-forward neural networks model. We consider both binary and multi-class classification. We develop a feed-forward neural networks model with embedding layers for multi-class classification, wherein a range of attack categories are treated as separate classes in addition to the class of normal traffic. We then extract weights from the embedding layers to encode high-dimensional categorical feature columns and build a second feed-forward neural networks model to perform binary classification. Details of the model architecture and tuning process are discussed in Section~\ref{setup}.

\item \textbf{Classification:} the final phase is to evaluate the classifiers on the specified testing set. The binary classifier outputs either a \emph{normal} or \emph{attack} label for each data instance in the testing set while the multi-class classifier outputs class labels (i.e., normal or specific attack class that it belongs to). Analysis of the evaluation results is presented in Section~\ref{analysis}.
\end{itemize}

\section{Evaluation and Analysis}

In this section, we first introduce the dataset used in the evaluation and then discuss feature extraction and feature preprocessing. Lastly, we provide description of the experiment setup and analysis of results.

\subsection{Dataset Description}

We use the BoT-IoT dataset developed by Koroniotis et al. \cite{Koroniotis2018arxiv} in the Cyber Range Lab of the UNSW Canberra Cyber Centre. Authors deployed a testbed to collect legitimate traffic from simulated IoT devices along with malicious traffic of various attacks. Five types of IoT devices were considered in the testbed to form a smart home scenario, including a weather station, a smart fridge, a smart thermostat, a remotely controlled garage door, and smart lights. In specific, the testbed consists of virtual machines (VMs) as attack or target machines, network devices (e.g., pfSense firewall), and simulated IoT devices running on normal VMs which are also connected to the IoT hub of the Amazon Web Services (AWS). These simulated IoT devices publish data using the MQTT procotol to the MQTT broker. There are four categories (ten subcategories) of malicious traffic generated in the dataset, including DoS and DDoS attacks over TCP, UDP and HTTP, probing attacks (port scanning and OS fingerprinting), and information theft (data exfiltration and keystroke logging). The detailed description of the testbed setup and statistics of attacks could be found in the paper \cite{Koroniotis2018arxiv}.

Reasons to adopt the dataset include the configuration of real testbed, collection of realistic attack traffic, generation of simulated IoT traffic in a smart home scenario, and inclusion of labelled data. 

\subsection{Feature Extraction}
\label{feature_extract}

Raw traffic in the BoT-IoT dataset was collected in the format of PCAP files. Authors labelled the raw traffic while collecting it and generated network flows by using the Argus tool. The processed traffic was stored in Argus files and further converted to CSV files. Labels in both Argus files and CSV files are flow-based. However, we need to retrieve labels on individual packets in order to perform the packet-level detection via our framework. We started with the collection of the PCAP files and obtained a total of 344 PCAP files from the BoT-IoT dataset. The PCAP files were organised into ten attack subcategories, namely DoS/DDoS over TCP, UDP and HTTP, service scan, OS fingerprinting, Keystroke logging, and data exfiltration.

We extracted individual packet header information from the PCAP files using the TShark tool, labelled the packets and output them into the CSV files. We then converted the CSV files to a single Apache Parquet file, unified the schema, and extracted roughly 2\% of the processed dataset. We explain details of each step in the following. 

\textbf{Feature description:} we only considered IP packets and extracted a total of 29 packet header fields from the PCAP files shown as follows. ARP packets were excluded because they are used to convert an IP address to a MAC address and irrelevant to the proposed attacks in the dataset. 

\begin{itemize}
\item Frame related fields: frame.time\_epoch, frame.len
\item IP related fields: ip.proto, ip.src, ip.dst, ipv6.src, ipv6.dst, ip.ttl, ip.id, ip.hdr\_len, ip.len, ip.flags.df
\item TCP related fields: tcp.srcport, tcp.dstport, tcp.stream, tcp.time\_delta, tcp.time\_relative, tcp.analysis.initial\_rtt, tcp.flags, tcp.window\_size\_value, tcp.hdr\_len, tcp.len
\item UDP related fields: udp.srcport, udp.dstport, udp.stream, udp.length 
\item HTTP related fields: http.response.code, http.request.method, http.content\_length
\end{itemize}

The TCP/IP protocol suite is vulnerable to a variety of attacks. The header fields of an individual packet could reflect generic features of the traffic and any deviation of values from a normal range in the fields could potentially become an indicator of attacks. For example, TCP, UDP, and HTTP are commonly exploited by attackers to launch protocol-based DDoS/DoS attacks; TCP and UDP are generally the protocols used in port scanning; OS fingerprinting works by sending TCP, UDP, and ICMP probes to known open and closed ports of the target; during post-exploitation, data could be exfiltrated via different channels (e.g., HTTP, TCP) while keystrokes could be observed via a reverse HTTP/TCP connection.

\textbf{Label mapping:} we analysed network flows in the original labelled dataset and replicated same labels for all individual packets within network flows. In specific, as described in~\cite{Koroniotis2018arxiv}, attack traffic of different subcategories was launched at different times with normal traffic generated in the background, which makes it easier to differentiate attacks; in addition, there are four bot machines deployed to launch attacks, which enables us to differentiate normal and attack traffic based on IP address of bot machines. Therefore, we developed a script to match the IP address of the bot machine and label packets with each attack subcategory. Labelled packets are stored in the CSV files for further processing. These CSV files with labelled traffic are placed in different folders based on the subcategory of attacks.

\textbf{File conversion and schema unification:} we converted the CSV files to a single Apache Parquet file, which reduces the file size and improves the processing speed. During conversion, we also unified the schema of all files. In specific, some ICMP packets contain the UDP packet inside the ICMP payload due to the recognition of the embedded packet by the network protocol analyser. Therefore, some IP related fields (e.g., ip.proto, ip.ttl) could be either string type (i.e., two values separated by comma) or double type. We enforced all values in those columns to be string type for consistency. Table~\ref{tb_dataset} shows statistics of the processed dataset with extracted features. 

\begin{table}[htb] 
\caption{Statistics of the processed dataset with extracted features.} 
\label{tb_dataset}
\centering
\begin{tabular}{|c|c|c|}
\hline
Category & Subcategory & Number of packets \\
\hline
\multirow{3}{*}{DDoS} & HTTP & 194417 \\
\cline{2-3} 
& TCP & 95643746 \\
\cline{2-3} 
& UDP & 174344218 \\
\hline 
\multirow{3}{*}{DoS} & HTTP & 287480 \\
\cline{2-3}
& TCP & 59977444 \\
\cline{2-3}
& UDP & 191208992 \\
\hline
\multirow{2}{*}{Reconnaissance} & OS fingerprinting & 827940 \\
\cline{2-3}
& Service scanning & 3814730 \\
\hline
\multirow{2}{*}{Information theft} & Data exfiltration & 301710 \\
\cline{2-3}
& Keylogging & 11387 \\
\hline
Normal & Normal & 23175520 \\
\hline
\multicolumn{2}{|c|}{Total} & 549787584 \\
\hline
\end{tabular}
\end{table}

\textbf{Dataset extraction:} we selected roughly 2\% of the processed dataset with extracted features. For each subcategory of attacks, we used all packets when the number of packets is lower than 1 million and selected 1 million packets if the number of packets is larger than 1 million. For the normal traffic, we selected roughly 2\% from all normal packets. In specific, for each attack subcategory with more than 1 million packets, the random selection was implemented through choosing a fraction of total packets of that attack subcategory (i.e., the fraction is 1 million divided by the total number of packets). Therefore, the number of chosen packets fluctuates slightly either above or below 1 million. The number of packets extracted is 11252406 including a mix of normal and attack traffic.

\subsection{Feature Preprocessing} 
\label{feature_preprocess}

We used a high performance computing (HPC) cluster with one Tesla V100 GPU and 93G RAM to preprocess extracted features in the Parquet and run the proposed deep learning model. We first selected fields that can be used as inputs into the deep neural networks model. We dropped the timestamp column (i.e., frame.time\_epoch) and IPv4/IPv6 address related columns (i.e., ip.src, ip.dst, ipv6.src, and ipv6.dst). We combined four pairs of columns respectively because values in those two columns of each pair do not overlap with each other. The four pairs of these columns include: tcp.srcport and udp.srcport; tcp.dstport and upd.dstport; tcp.len and udp.length; tcp.stream and udp.stream. Combined columns are named as src.port, dst.port, length, and stream. We split columns with comma separated values (i.e., ip.ttl, ip.id, ip.hdr\_len, and ip.len) into two columns while non-embedded packets are filled with NaN values in the second column. Afterwards, we filled NaN/None values with either numeric values or string values. For numeric columns with meaningful zero values (e.g., ip.flags.df, tcp.flags, src.port, and dst.port), we obtained the maximum value in each column and filled NaN values with the maximum value plus 1. For the string column (i.e., http.request.method), we filled None values with `0' (in string type). For other columns, we filled NaN values with 0. 

After dropping several irrelevant columns, duplicated rows were introduced. We dropped redundant rows because the testing traffic should not have duplicates from the training traffic. Without redundancy, we will be able to check how the model will perform on unseen data in the testing phase. Table~\ref{tb_attack_stats} presents statistics of the dataset before and after dropping duplicates rows. We can see that the total number of packets being kept is roughly 81.44\% of the processed dataset while the majority of packets being dropped are normal packets with a drop rate of 45.08\%.

\begin{table}[htb] 
\caption{Statistics of the dataset before and after dropping duplicated rows.} 
\label{tb_attack_stats}
\centering
\begin{tabular}{|c|c|c|c|}
\hline
\multirow{2}{*}{Category} & \multirow{2}{*}{Subcategory} & \multicolumn{2}{c|}{Drop duplicated rows}\\
\cline{3-4}
&  & Before & After\\
\hline
\multirow{3}{*}{DDoS} & HTTP & 194417 & 194417 \\
\cline{2-4} 
& TCP & 999444 & 999444 \\
\cline{2-4} 
& UDP & 1000055 & 1000054 \\
\hline 
\multirow{3}{*}{DoS} & HTTP & 287480 & 287480 \\
\cline{2-4}
& TCP & 998703 & 998703 \\
\cline{2-4}
& UDP & 999977 & 999977 \\
\hline
\multirow{2}{*}{Reconnaissance} & OS fingerprinting & 827940 & 827058 \\
\cline{2-4}
& Service scanning & 999895 & 999895 \\
\hline
\multirow{2}{*}{Information theft} & Data exfiltration & 301710 & 301710 \\
\cline{2-4}
& Keylogging & 11387 & 11387 \\
\hline
Normal & Normal & 4631398 & 2543626 \\
\hline
\multicolumn{2}{|c|}{Total} & 11252406 & 9163751 \\
\hline
\end{tabular}
\end{table}

We identified feature columns with categorical data. For columns with low-dimensional categorical variables (i.e., ip.proto, tcp.flags, ip.flags.df, and http.response.code), we applied one-hot encoding and dropped original columns. We refer these columns and other columns with non-categorical values as input columns. For columns with highly-dimensional categorical variables (i.e., src.port, dst.port, and http.request.method), we separated them from other columns and will apply embedding in the deep neural networks model to compute a dense representation of ports and HTTP request methods. We refer these three columns as embedding columns. We normalised input columns within a given range between (-1, 1) to avoid extreme values and potentially help in speeding up calculations. We stored input columns, port embedding columns and embedding column for HTTP request methods as three arrays. 

We also encoded label columns. We considered both binary classification (i.e., classify normal packets and attack packets from each subcategory of attacks separately) and multi-class classification (i.e., classify normal packets and malicious packets from DDoS/DoS, information theft and reconnaissance attacks). In the binary classification, a normal packet is labelled as 0 while an attack packet is labelled as 1; in the multi-class classification, packets from each class are labelled from 0 to 3 respectively. Labels for each classification were stored as an array.

\subsection{Experimental Setup}
\label{setup}

We used TensorFlow library and Keras. All our experiments were effected on the HPC cluster with Tesla V100 GPU. We split the extracted dataset with 64\%-16\%-20\% on training, validation, and testing in a stratified fashion thus training, validation, and testing sets have approximately the same percentage of samples of each class as the complete set. We introduced class weights to the training data due to the imbalanced samples of different classes. The weight for each class was given by the inverse of the quotient value from dividing the packet count in one particular class by the maximum packet count among all classes. In this manner, the under-represented class with a relatively smaller number of samples obtains a higher weight value. 

\subsubsection{Model Design}

We consider to use feed-forward neural networks (FNN) models. For multi-class classification, the FNN model has two embedding layers and three dense layers. For binary classification, we applied the concept of transfer learning by using the embedding vectors resulting from embedding layers in the FNN model for multi-class classification. We initialised both models with random weights. We applied Adam optimiser and a sparse categorical cross-entropy loss function. We denote the FNN model for multi-class classification as mFNN and the model for binary classification as bFNN and explain them in details as follows.

\begin{figure}[h!]
  \centering
 \includegraphics[width=0.9\textwidth]{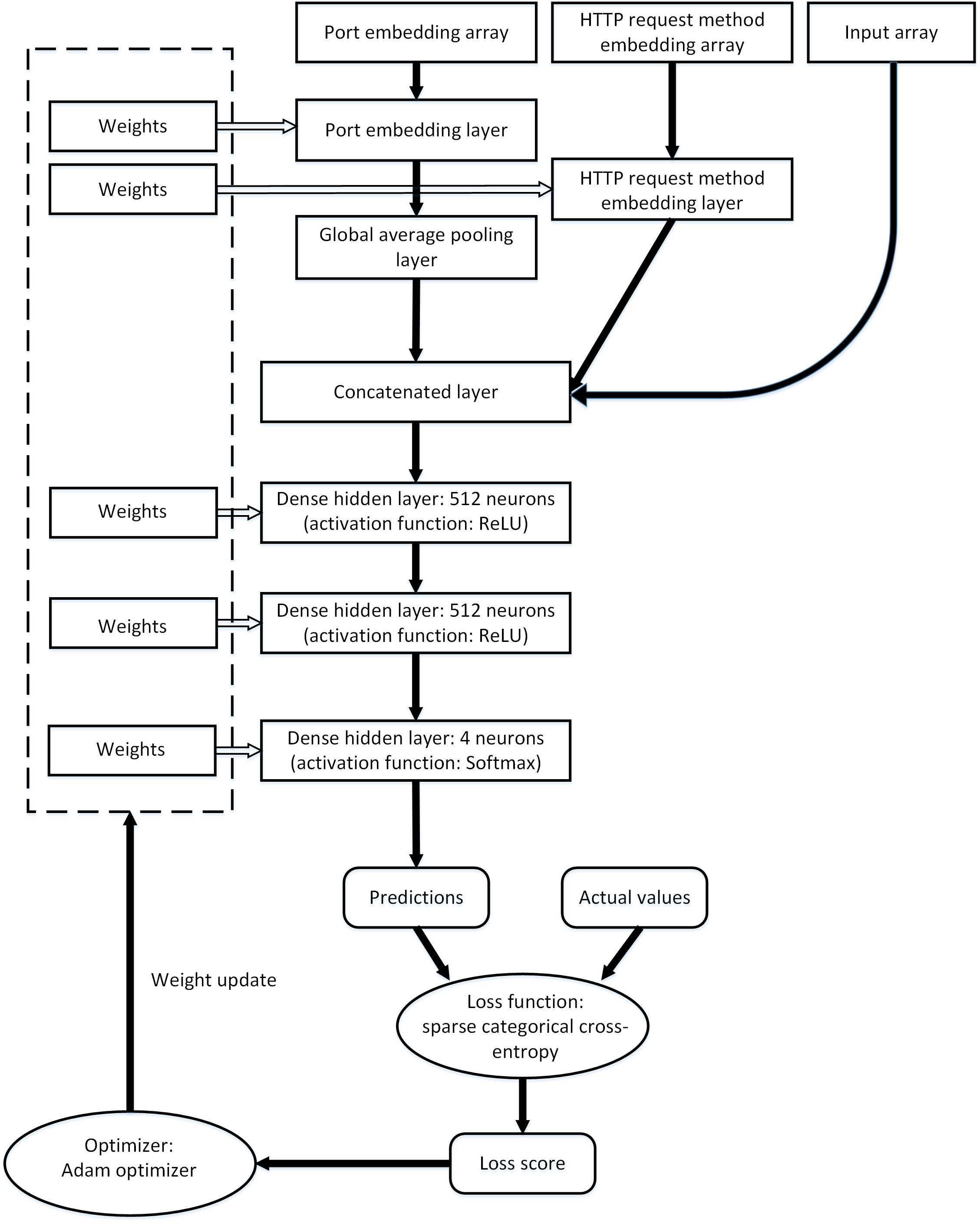}
\caption{FNN model for multi-class classification.}
\label{fig_mfnn}
\end{figure}

\textbf{mFNN:} Figure~\ref{fig_mfnn} shows the architecture of the FNN model along with interactions with the loss function and the solver for multi-class classification. mFNN model takes three arrays as inputs in which two of them are inputs for the embedding layers. The port embedding layer has an input dimension of 65537 (with all official and non-official ports between 0 and 65535 and 65536 encoded for NaN values) and an output dimension of 16 while HTTP request method embedding layer has an input dimension of 93 (with a number of 92 official request methods and a string encoded for none values) and an output dimension of 4. The model consists of three dense hidden layers with each input neuron being connected with each output neuron in each hidden layer. The first two hidden layers have 512 neurons and use a ReLU activation function in each layer while the last hidden layer has 4 neurons corresponding to 4 classes and uses a Softmax function which turns numbers into probabilities that sum to one. 

\begin{figure}[h!]
  \centering
 \includegraphics[width=0.65\textwidth]{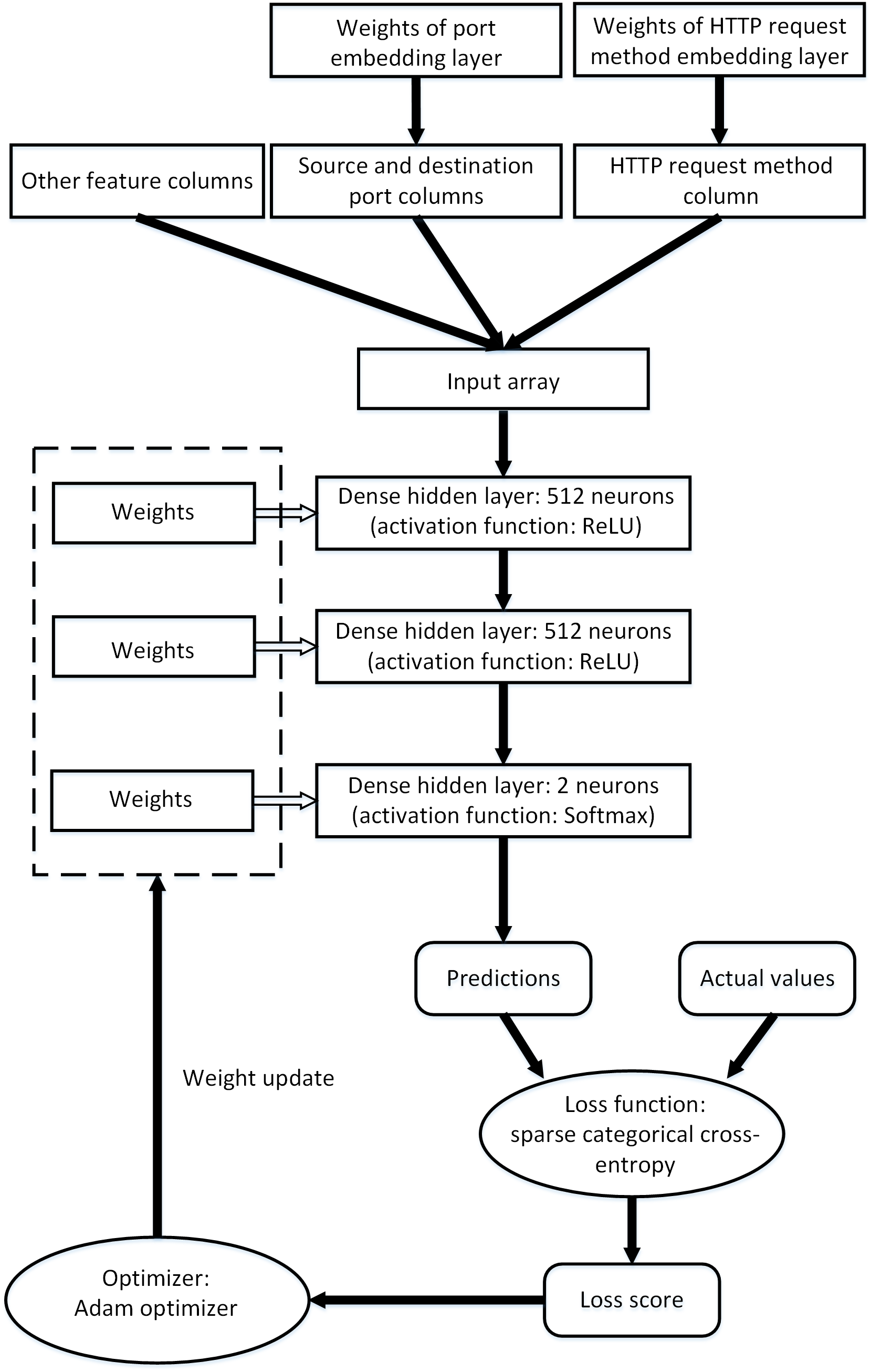}
\caption{FNN model for binary classification.}
\label{fig_bfnn}
\end{figure}

\textbf{bFNN:} Figure~\ref{fig_bfnn} shows the architecture of the FNN model for binary classification. bFNN model uses weights of the two embedding layers from mFNN model to encode source/destination ports and HTTP request methods; the encoded columns (i.e., src.port, dst.port, and http.request.method) are then combined with other columns and taken as input into the model. bFNN model has three dense hidden layers with the first two hidden layers consisting of 512 neurons and using a ReLU function in each layer and the last hidden layer having 2 neurons and a Softmax function.

\subsubsection{Model Tuning}

We tuned mFNN model with different hyperparameters and three regularisation techniques. We used same hyperparameters and regularisation optimisation for bFNN model because we want to apply the generalisation of the model to different cases. We discuss the regularisation and tuning procedure for the multi-class classification in details in the following paragraphs. We used the FNN model with three dense hidden layers, 512 neurons in the first two hidden layers and default learning rate of Adam optimiser (i.e., 0.001) as the baseline model. 

\textbf{Regularisation:} we first experimented with three commonly used regularisation techniques including L1, L2, and dropout. Results show the classification accuracy does not improve. That said, regularisation is proven effective for preventing over-fitting. Since our model is trained on a dataset with millions of samples, the over-fitting problem is a less concern.

\textbf{Hyperparameter tuning:} we then experimented with different architectures of the neural networks, including adding/reducing hidden layers (ranging from two to four hidden layers), increasing the number of neurons in hidden layers (including 512 or 1024 neurons; excluding the last dense layer), and adding dropout with dropout rate of 0.1 in hidden layers (excluding the last dense layer). Results demonstrate that the baseline neural networks model with three hidden layers and 512 neurons in the first two layers has the best performance. Different variants of the neural networks model do not increase the classification accuracy. One possible reason could be that the complexity of the model is either too small or too large. When the model complexity is too small, the model may not be capable of capturing the variation of different classes in the training data; when the complexity is too large, the model tends to over-fit different classes in the training data. Therefore, the model with too large/small complexity would make the performance on specific classification problems worse. Lastly, we experimented with different learning rates of Adam optimiser, including 0.01, 0.001, 0.0001, and 0.00001. We chose the learning rate of 0.0001 which gives highest accuracy. A larger learning rate tends to oscillate over training epochs, causing weights to diverge while a smaller learning rate could possibly get stuck on a sub-optimal solution.

We also applied an early stopping technique with a patience of 5 iterations to terminate the training as soon as the validation loss reaches a minimum. This has the advantage of reducing over-fitting problem. The FNN model was trained with a batch size of 256 and the maximum epoch was set to be 20. 

\subsection{Analysis of Results}
\label{analysis}

We discuss results for the multi-class classification and binary classification. We compute confusion matrices for both classification problems and use following metrics to evaluate the model performance in the binary classification: accuracy, precision, recall, and F1 score. In specific, accuracy is defined as the number of correct predictions, including true positive (TP) and true negative (TN) predictions, divided by the total number of predictions; precision is the number of TP predictions divided by the total number of predicted positive class values; recall is the number of TF predictions divided by the number of actual positive class values; F1 score is weighted average of precision and recall. A low precision indicates a large number of false positive (FP) predictions (normal packets classified as attack packets) while a low recall could indicate a large number of false negative (FN) predictions (attack packets classified as normal packets); a good F1 score indicates low FP/FN predictions because it combines precision and recall. Table~\ref{tb_metrics} shows definitions of the metrics in terms of positives and negatives.

\begin{table}[htb]
\caption{Definitions of metrics.}
\label{tb_metrics}
\centering
\renewcommand{\arraystretch}{2.0}
\begin{tabular}{|c|c|}
\hline
Accuracy & $\dfrac{TP+TN}{TP+TN+FP+FN}$ \\
\hline
Precision & $\dfrac{TP}{TP+FP}$ \\
\hline
Recall & $\dfrac{TP}{TP+FN}$ \\
\hline
F1 score & $2\times \dfrac{Precision\times{Recall}}{Precision+Recall}$ \\
\hline
\end{tabular}
\end{table}

\textbf{Multi-class classification:} Table~\ref{tb_multi} shows the confusion matrix for the multi-class classification with packet count and percentage value. mFNN model achieves a high accuracy of 99.79\% over four classes of traffic classification. For predicting traffic of one class over traffic of the other three classes, the accuracy is above 99\% respectively. In addition, there is no misclassification between DDoS/DoS or normal traffic and information theft traffic. For DDoS/DoS attacks, the model misclassified 2314 packets (0.258\%) as reconnaissance traffic while for reconnaissance attacks, the model misclassified 1540 packets (0.421\%) as DDoS/DoS traffic. The reason could be behaviours of reconnaissance attacks reflected through the current feature set have similarity with behaviours of DDoS/DoS attacks which makes it difficult to distinguish between these two attacks. Regarding the runtime, the mFNN model completed the training and validation procedures with roughly 42 minutes. 

\begin{table*}[h!]
\caption {Confusion matrix for the multi-class classification for various attack types} \label{tb_multi} 
 \centering
 	\resizebox{\textwidth}{!}{%
\begin{tabular}{ |c|c|c|c|c| } 
\hline
Actual/Predicted & Normal & DDoS/DoS & Information theft & Reconnaissance \\
\hline
Normal & 508702 \textbf{(99.995\%)} &  1 & 0 & 22 \\
\hline
DDoS/DoS & 4 & 893697 \textbf{(99.741\%)} & 0 & 2314 (0.258\%) \\
\hline
Information theft & 0 & 0 & 62615 \textbf{(99.992\%)} & 5 \\
\hline
Reconnaissance &  13 & 1540 (0.421\%) & 13 & 363825 \textbf{(99.571\%)} \\
\hline
\end{tabular}}
\end{table*}

\begin{table*}[h!]

\caption {Confusion matrices for the binary classification for various attack types} \label{tb_binary} 
 \centering

\begin{tabular}{ c c }
\begin{tabular}{ |p{.25\textwidth}|p{.15\textwidth}|p{.25\textwidth}|} 
\hline
Actual/Predicted & Normal & DDoS over HTTP \\
\hline
Normal & 508725 & 1 \\
\hline
DDoS over HTTP & 0 & 38883 \\
\hline
\end{tabular} \\
\\
\begin{tabular}{ |p{.25\textwidth}|p{.15\textwidth}|p{.25\textwidth}|} 
\hline
Actual / Predicted & Normal & DDoS over TCP \\
\hline
Normal & 508724 & 1 \\
\hline
DDoS over TCP & 0 & 199889 \\
\hline
\end{tabular} \\
\\
\begin{tabular}{ |p{.25\textwidth}|p{.15\textwidth}|p{.25\textwidth}|} 
\hline
Actual / Predicted & Normal & DDoS over UDP \\
\hline
Normal & 508725 & 0 \\
\hline
DDoS over UDP & 0 & 200011 \\
\hline
\end{tabular} \\
\\
\begin{tabular}{ |p{.25\textwidth}|p{.15\textwidth}|p{.25\textwidth}|} 
\hline
Actual / Predicted & Normal & DoS over HTTP \\
\hline
Normal & 508725 & 1 \\
\hline
DoS over HTTP & 0 & 57496 \\
\hline
\end{tabular} \\
\\
\begin{tabular}{ |p{.25\textwidth}|p{.15\textwidth}|p{.25\textwidth}|} 
\hline
Actual / Predicted & Normal & DoS over TCP \\
\hline
Normal & 508724 & 1 \\
\hline
DoS over TCP & 1 & 199740 \\
\hline
\end{tabular}\\
\\
\begin{tabular}{ |p{.25\textwidth}|p{.15\textwidth}|p{.25\textwidth}|} 
\hline
Actual / Predicted & Normal & DoS over UDP \\
\hline
Normal & 508725 & 0 \\
\hline
DoS over UDP & 0 & 199996 \\
\hline
\end{tabular} \\
\\
\begin{tabular}{ |p{.25\textwidth}|p{.15\textwidth}|p{.25\textwidth}|} 
\hline
Actual / Predicted & Normal & Data exfiltration \\
\hline
Normal & 508726 & 0 \\
\hline
Data exfiltration & 0 & 60342 \\
\hline
\end{tabular}\\
\\
\begin{tabular}{ |p{.25\textwidth}|p{.15\textwidth}|p{.25\textwidth}|} 
\hline
Actual / Predicted & Normal & Keylogging \\
\hline
Normal & 508726 & 0 \\
\hline
Keylogging & 0 & 2277 \\
\hline
\end{tabular}\\
\\
\begin{tabular}{ |p{.25\textwidth}|p{.15\textwidth}|p{.25\textwidth}|} 
\hline
Actual / Predicted & Normal & Service scan \\
\hline
Normal & 508710 & 16 \\
\hline
Service scan & 4 & 199975 \\
\hline
\end{tabular}\\
\\
\begin{tabular}{ |p{.25\textwidth}|p{.15\textwidth}|p{.25\textwidth}|} 
\hline
Actual / Predicted & Normal & OS fingerprint \\
\hline
Normal & 508721 & 4 \\
\hline
OS fingerprint & 0 & 165412 \\
\hline
\end{tabular}
\end{tabular}
\end{table*}

\textbf{Binary classification:} Table~\ref{tb_binary} shows confusion matrices for the binary classification. We can see the classification for normal traffic and malicious traffic from each subcategory of attacks achieves high accuracy, precision, recall, and F1 score (i.e., 99.99\%) with only a few packets misclassified. 

All attack traffic in the dataset was generated via standard tools (e.g., Nmap, Hping3, Metasploit framework). In reality, attackers could use different options of the same tools (e.g., packet sending rate, payload size) or different tools for the attack configuration; some intelligent attackers could also apply evasion techniques. However, the header fields contained in the packets could be kept similar to achieve the same attack goal (e.g., a SYN flood attack needs to set SYN flag in tcp.flags field). Therefore, the attack traffic in the dataset still reflects genuine behaviours by attackers, which demonstrates the applicability of the proposed model. 

\subsection{Comparison of Results}

As a final step, we applied the SVM model~\cite{RAJKUMAR20111328} to the multi-class classification problem and compared the classification accuracy and runtime of SVM with those of mFNN model. We discuss the input, model settings and results for multi-class classification in detail in the following paragraphs. 

We used the same dataset without redundant row values shown in Table~\ref{tb_attack_stats}. We identified categorical columns and mapped three highly-dimensional categorical variables (i.e., src.port, dst.port, and http.request.method) to low-dimensional categorical variables. For src.port and dst.port columns, we obtained a list of common ports (i.e., 20, 21, 22, 23, 25, 42, 43, 53, 80, 161, 443), kept these common ports in original port columns and map uncommon ports to new columns. This list was constructed based on well-known ports for various services, including FTP, SSH, Telnet, DNS, SMTP, HTTP/HTTPS, and SNMP. We generated two new columns for src.port and dst.port respectively: one for the port number between the range of 0 and 1023 but does not belong to the well-known port list; another for the port number larger than 1023 (i.e., registered/private ports). For http.request.method column, we maintained row values containing a single method (i.e., `POST', `GET', `OPTIONS', `PROPFIND', `HEAD', and `TRACE') or `0' for NaN and mapped values containing comma separated multiple methods to a string named `MULTIPLE' (e.g., `GET,GET', `HEAD,HEAD,HEAD'). We then applied one-hot encoding to all low-dimensional categorical variables. Due to the long convergence time of the SVM model, we performed random subsampling out of 9163751 samples with sample sizes of 10000, 100000, and 1000000 (i.e., 10.9\%). We randomly selected samples so as to follow the same distribution (i.e., proportion) of data from different classes. We also applied feature scaling to normalise inputs within a range between (-1, 1).

We used a linear Support Vector Classifier (SVC) from the Scikit Learn library\footnote{Scikit is widely accessible at \url{https://scikit-learn.org}}. Due to the imbalanced nature of the different classes in the dataset, class weights were introduced to the training data. We followed the same weight calculation performed for mFNN model. For each subsampled set, we tuned the SVC via grid search by cross validation with a list of penalty parameter values, including 0.1, 1, 10, 100, 1000, and 10000. The penalty parameter controls the trade-off between smooth decision boundary and correct classification of training points; a higher value allows fewer outliers while a lower value allows more outliers thus avoiding over-fitting. We applied a 5-fold cross validation on each set with a 80\%-20\% training and testing split. We placed a limit of 10000 on maximum number of iterations to be run. In line with our other experiments, we used the same HPC cluster. We retrieved the optimal penalty parameter value from the grid search for each set (i.e., 1000 for the set with sample size of 10000 and 1000000; 100 for the set with sample size of 100000) and applied the SVC with the optimal value to training and testing.

Table~\ref{tb_svc} illustrates accuracy and runtime results for the three subsampled sets. The class size represents the number of selected samples in each class in the order of normal, DDoS/DoS, information theft, and reconnaissance traffic while the runtime is the training time under the optimal penalty parameter value. From the table, we can see that the SVC yields a slightly higher accuracy with larger sample size. That said, the runtime also increases moderately. In addition, the model did not converge due to the limitation on maximum number of iterations. So without the iteration limit, the model could run until it converges thus incurring higher accuracy; however, this may increase the runtime dramatically due to the uncertainty of the convergence time. Thus, we could conclude the proposed FNN model for the multi-class classification is more time efficient over large dataset compared with the SVC.

\begin{table*}[h!]
\caption {Accuracy and runtime of SVC for the multi-class classification} \label{tb_svc} 
 \centering
\begin{tabular}{ |c|c|c|c|c| } 
\hline
Sample size & Class size & Accuracy & Runtime (min)\\
\hline
10K & (2817, 4912, 328, 1943) & 97.90\% & 0.35 \\
\hline
100K & (27739, 49203, 3418, 19640) & 98.15\% & 8.40 \\
\hline
1M & (277800, 488896, 34037, 199267) & 98.26\% & 144.61 \\
\hline
\end{tabular}
\end{table*}

\section{Discussion and Future Work}

By extracting the header field information of individual packets as generic features and developing the FNN models, we are able to differentiate normal and attack traffic with high accuracy and low FP and FN predictions. However, we have encountered several issues and aim to address these in our future work.

For the multi-class classification, we considered four categories at the current stage because we observed the model yielded sub-par performance for certain subcategories of attacks (e.g., DDoS over HTTP and DoS over HTTP). The reason could be that the field information for the individual packet could not capture certain attack behaviour in a large scale (e.g., a flood of traffic generated in DDoS attacks). Therefore, we are planning to develop a classifier to differentiate each subcategory of attacks as part of the future work. We will include the timestamp information, process header fields of individual packets aggregated from timestamps, and develop a new deep neural networks model to incorporate these new features. Long-short term memory networks could be the best option to model time series data. In addition, the current classifier works in the batch mode. We will implement the online mode for real time detection. We also plan to investigate feature ranking techniques for time-series feature-based classifiers. 

\section{Conclusion}

Deep learning techniques have shown its potential to serve as a solution for the intrusion detection problem in IoT networks. Through the proposal of an intelligent intrusion detection approach, we have developed feed-forward neural networks models along with the extraction and preprocessing of the header field information in individual packets, treated as generic features, and adopted the concepts of neural network embedding and transfer learning to obtain encoding of highly-dimensional categorical features of data. We have tested the efficacy of the models for both binary and multi-class classification on a dataset comprising realistic network traffic. Through the evaluation results, we have demonstrated the performance of the proposed approach. In particular, the capability of the classifier in binary classification was illustrated through results close to 99.99\% across all evaluation measures including accuracy, precision, recall, and F1 score while the detection accuracy of approximately 99.79\% was achieved for multi-class classification. The results are significant and warrant further research in this domain of cyber security for IoT networks.

\bibliographystyle{elsarticle-harv}
\bibliography{jrn-ext}

\end{document}